\documentclass[
 pra,
 10pt,
 a4paper,
 superscriptaddress,
 nofootinbib,
 twocolumn,
 floatfix,
 longbibliography,
 showkeys,
 floatfix
]{revtex4-2}
\usepackage{soul}
\usepackage[T1]{fontenc}
\usepackage{float}
\usepackage[american]{babel}
\usepackage[utf8x]{inputenc}
\usepackage{grffile} 
\usepackage{newtxtext,newtxmath}
\usepackage{microtype}
\usepackage{amsmath}
\usepackage{bbm}
\usepackage{bm}
\usepackage{mathtools}
\usepackage{physics}
\usepackage{dsfont}
\usepackage{siunitx} 

\usepackage[section]{placeins}
\usepackage[cal=cm,scr=dutchcal]{mathalpha}
\newcommand{\I}{\mathrm{i}}
\newcommand{\E}{\mathrm{e}}

\DeclareFontFamily{OMS}{oasy}{\skewchar\font48 }
\DeclareFontShape{OMS}{oasy}{m}{n}{
         <-5.5> oasy5     <5.5-6.5> oasy6
      <6.5-7.5> oasy7     <7.5-8.5> oasy8
      <8.5-9.5> oasy9     <9.5->  oasy10
      }{}
\DeclareFontShape{OMS}{oasy}{b}{n}{
       <-6> oabsy5
      <6-8> oabsy7
      <8->  oabsy10
      }{}
\DeclareSymbolFont{oasy}{OMS}{oasy}{m}{n}
\SetSymbolFont{oasy}{bold}{OMS}{oasy}{b}{n}

\DeclareMathSymbol{\smallleftarrow}     {\mathrel}{oasy}{"20}
\DeclareMathSymbol{\smallrightarrow}    {\mathrel}{oasy}{"21}
\DeclareMathSymbol{\smallleftrightarrow}{\mathrel}{oasy}{"24}

\newcommand{\D}{\mathrm{d}}

\NewDocumentCommand{\Ds}{sm}{
  \IfBooleanTF{#1}
   {
    \mathrm{d}#2\,
   }
   {
    \mathop{}\!\mathrm{d}#2
   }
}

\NewDocumentCommand{\FuncDs}{sm}{
  \IfBooleanTF{#1}
   {
    \mathscr{D}#2\,
   }
   {
    \mathop{}\!\mathscr{d}#2
   }
}

\newcommand{\Vect}[1]{\boldsymbol{#1}}

\NewDocumentCommand{\intRlap}{e{_^}}{
  \DOTSI
  \int_{\IfValueT{#1}{\mathrlap{#1}}}^{\IfValueT{#2}{\mathrlap{#2}}}
  \mspace{-4mu}
}

\newcommand{\PhaseSpaceX}{\chi}
\newcommand{\PhaseSpaceP}{\wp}
\newcommand{\PhaseSpaceXVec}{\Vect{\chi}}
\newcommand{\PhaseSpacePVec}{\Vect{\wp}}
\newcommand{\PhaseSpaceAction}{\mathcal{A}}
\NewDocumentCommand{\PhaseSpaceArgs}{ O{} O{}}{\PhaseSpaceX^{#1}_{#2},\PhaseSpaceP^{#1}_{#2},\PhaseSpaceAction^{#1}_{#2}}
\NewDocumentCommand{\PhaseSpaceArgsVec}{ O{} O{}}{\PhaseSpaceXVec^{#1}_{#2},\PhaseSpacePVec^{#1}_{#2},\PhaseSpaceAction^{#1}_{#2}}

\makeatletter
\newcommand{\xsize}[1]{\bBigg@{#1}}
\newcommand{\vast}{\bBigg@{4}}
\newcommand{\Vast}{\bBigg@{5}}
\makeatother

\makeatletter
\newcommand{\Eqlfill@}{\arrowfill@\Relbar\Relbar\Relbar}
\newcommand{\eql}[2][]{\ext@arrow 0066\Eqlfill@{#1}{#2}}
\makeatother

\makeatletter
\def\leftharpoonfill@{\scriptsize\bf\bf\arrowfill@\leftharpoonup\relbar\relbar}
\def\rightharpoonfill@{\scriptsize\bf\bf\arrowfill@\relbar\relbar\rightharpoonup}
\def\overarrow@#1#2#3{\vbox{\ialign{##\crcr#1#2\crcr
 \noalign{\nointerlineskip}$\m@th\hfil#2#3\hfil$\crcr}}}
\newcommand{\overrightharpoon}{
  \mathpalette{\overarrow@\rightharpoonfill@}}
\newcommand{\overleftharpoon}{
  \mathpalette{\overarrow@\leftharpoonfill@}}
\makeatother
\usepackage{graphicx}
\usepackage[svgnames]{xcolor}

\definecolor{cset-aps-blueberry}{RGB}{28,128,158}
\definecolor{cset-aps-blue}{RGB}{46,44,184}
\definecolor{cset-aps-turquoise}{RGB}{0,67,88}
\definecolor{cset-aps-limegreen}{RGB}{190,219,67}
\definecolor{cset-aps-green}{RGB}{31,138,112}
\definecolor{cset-aps-yellow}{RGB}{255,225,25}
\definecolor{cset-aps-orange}{RGB}{253,116,0}
\definecolor{cset-aps-red}{RGB}{219,0,43}

\definecolor{cset-aps-kobalt-medium}{RGB}{62,54,222}
\definecolor{cset-aps-kobalt-dark}{RGB}{28,24,150}

\definecolor{cset-aps-my-label-red}{RGB}{202,0,17}
\definecolor{cset-aps-my-label-blue}{RGB}{53,71,140}
\definecolor{cset-aps-my-label-gray}{RGB}{145,145,145}

\usepackage{hyperref}
\hypersetup{
    colorlinks=true,
    linkcolor={cset-aps-red},
    linkbordercolor={cset-aps-red},
    filecolor={cset-aps-orange},
    filebordercolor={cset-aps-orange},
    citecolor={cset-aps-kobalt-medium},
    citebordercolor={cset-aps-kobalt-medium},
    urlcolor={cset-aps-kobalt-dark},
    urlbordercolor={cset-aps-kobalt-dark},
    menucolor={cset-aps-limegreen},
    menubordercolor={cset-aps-limegreen},
    breaklinks=true,
    pdfborderstyle={/S/U/W 2},
    pdfpagemode=UseOutlines,
    pdfstartpage={1},
}

\usepackage[capitalise]{cleveref}

\crefname{secinapp}{appendix}{appendices}
\Crefname{secinapp}{Appendix}{Appendices}

\usepackage[autostyle]{csquotes}

\usepackage{array}
\usepackage{tabularx}
\usepackage{booktabs}
\usepackage{multirow}

\newcommand{\affULM}{\address{Institut f{\"u}r Quantenphysik and Center for Integrated Quantum Science and Technology (IQST), Universit{\"a}t Ulm, Albert-Einstein-Allee 11, D-89069 Ulm, Germany}}

\newcommand{\orcid}[1]{\href{https://orcid.org/#1}{\includegraphics[width=7pt]{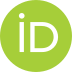}}}

\newcommand{\Kick}{\ensuremath{\hat{D}}}
\newcommand{\Kickdag}{\ensuremath{\hat{D}^{\dagger}}}
\renewcommand{\vec}{\mathbf}
\begin{document}

\title{Decoupling of external and internal dynamics in driven two-level systems
}
\collaboration{Published version available in  \href{https://doi.org/10.1103/PhysRevResearch.6.043153}{Phys. Rev. Res. \textbf{6}, 043153 (2024)};
licensed under a \href{http://creativecommons.org/licenses/by/4.0/}{Creative Commons Attribution [CC BY]} license.}

\author{Samuel Böhringer~\orcid{0009-0003-9835-4009}}
\affULM
\email{samuel.boehringer@uni-ulm.de}

\author{Alexander Friedrich~\orcid{0000-0003-0588-1989}}
\affULM

\begin{abstract}
\noindent We show how a laser driven two-level system including quantized external degrees of freedom for each state can be decoupled into a set of oscillator equations acting only on the external degrees of freedom with operator valued damping representing the detuning. 
We give a way of characterizing the solvability of this family of problems by appealing to a classical oscillator with time-dependent damping.
As a consequence of this classification we (a) obtain analytic and representation-free expressions for Rabi oscillations including external degrees of freedom with and without an external linear potential, (b) show that whenever the detuning operator can be diagonalized (analytically or numerically) the problem decomposes into a set of classical equations and (c) we can use the oscillator equations as a perturbative basis to describe Rabi oscillations in weak but otherwise arbitrary external potentials. 
Moreover, chirping of the driving fields phase emerges naturally as a means of compensating the Ehrenfest/mean-value part of the detuning operator's dynamics while the presence of driving phase noise leads to a stochastic evolution equation of Langevin type.
Lastly, our approach is representation free with respect to the external degrees of freedom and as consequence a suitable representation or basis expansion can be chosen a posteriori depending on the desired application at hand.
\end{abstract}
\maketitle

\section{Introduction}
The two-level system and the harmonic oscillator constitute the two foundational models upon whose understanding most of today's descriptions of quantum systems are built upon.
In quantum optics and cavity quantum electrodynamics the two-level atom in resonator combines both of these systems into a single, interacting system. In this form, this duo serves as the workhorse model of these fields, e.g. as the Jaynes-Cummings~\cite{Larson2021} or Rabi model.
Fundamentally, these models describe the simplest conceivable coupling between a spin system and a continuous degree of freedom.
As such, they have been studied extensively in quantum optics~\cite{Browne2017,Fabre2020} and feature heavily in applications ranging from atomic clocks~\cite{Ludlow2015,Vanier2015} over atom interferometers and inertial sensors~\cite{Cronin2009,Mehlstaubler2018,Bongs2019,Narducci2022} and atomtronics~\cite{Amico2022} to quantum repeaters~\cite{Azuma2023} and qubits~\cite{Kjaergaard2020} in the field of quantum information.

Many of these approaches,~experiments and systems involve a two-level system with motional degrees of freedom, driven coherently by an external (classical) field.
This combination leads to coherent Rabi oscillations~\cite{Shore2011} of the two-level system for resonant external driving fields. Without external, that is motional degrees of freedom, there even is a wide range of analytic solutions available~\cite{Demkov1969,Bambini1981,Hioe1985,Zakrzewski1985,Stehle1987,Vitanov2007} for different pulse envelopes in the driving field.
In qubit applications~\cite{Undseth2023,Mortezapour2018} and atomic clocks~\cite{Oelker2019Oct,McGrew2019Apr} the motional degrees of freedom of the system are often seen as parasitic effects limiting the ultimate precision and stability of such systems. In such cases the motional degrees of freedom might represent, e.g.~residual motion in the trap, stray fields or vibrations.
In contrast, in atom optics the coupling between internal dynamics and external motion is key in transferring momentum to atoms via diffraction and thus opens the door to perform atom interferometry in the first place~\cite{Colella1975,Kasevich1991Jul,Giese2015}.
In all of these cases the joint system, consisting of continuous external and spin-like internal degrees of freedom, is driven by the external field to facilitate a transition. Hence, a full description also necessitates solving the coupled Schrödinger dynamics of internal and external motion, as they are non-commutative in general.
Even in the simplest case of a two-level atom interacting with a potential in free space we are thus left with an intricately coupled and complex six dimensional partial differential equation system when all internal and external degrees of freedom are accounted for.
Although nowadays a feasible task, this is in general still numerically challenging.
As a consequence, finding simplifications for this problem is of immense interest as they may proof useful in many applications or lead to new insights and methods.
In this article we provide such a simplification by formally decoupling internal and external dynamics of a driven spin system under quite generic assumptions.

\subsection{Overview and Key Results}
In this article particularly, we show that internal spin- and external dynamics can be decoupled and the Schrödinger equation is equivalent to a decoupled system of driven oscillator equations with operator valued coefficients.
These oscillator equations and their solution give direct access to the matrix elements of the time-evolution operator with respect to the spin degrees of freedom of the system.
Furthermore, they can be solved directly for specific external potentials, among them the free particle and a linear potential, which leads us to representation free analytic expressions for the time-evolution operator in these cases.
Moreover, the oscillator equations and their solutions can serve as basis for a suitable perturbation expansion applicable to arbitrary but weak external potentials.
This perturbation expansion directly accounts for the oscillatory nature of the spin dynamics without relying on a specific choice of interaction picture first in contrast to typical applications of perturbation theory via the Dyson series.
Lastly, we apply our perturbation method to the case of a two-level atom in a quadratic potential and show how the combination of internal and external dynamics modifies the Rabi oscillations.

Our manuscript is organized as follows: Sec.~\ref{sec:decouple} demonstrates the process of decoupling the Rabi model, which incorporates external degrees of freedom in arbitrary potentials, into oscillator equations solely within the external space. This derivation is followed by three subsections discussing different aspects of the model:
In Sec.~\ref{sec:equivalent-form} we explicitly show how our oscillator equation can be brought into a (less favourable) form which coincides in special cases with a previously derived result by Marzlin~et al. in Ref.~\cite{Marzlin1996Feb} for the case of a linear potential. In Sec.~\ref{sec:resonance-condition-emerges} we discuss the emergence of the resonance condition from our operator approach. 
Following this in Sec.~\ref{sec:chirping-and-phase-noise} we elaborate how chirping the driving field can be used to on average remove the action of the potential and how phase noise from the driving field enters.

In the following Sec.~\ref{sec:applications} we apply our results to the cases of (a) a free particle and (b) a particle in a linear potential where we obtain an analytical solution in both cases by considering a classical oscillator equation.
Moreover, in ~Sec.~\ref{subsec:numerics} we derive a perturbative approach based on our oscillator equation for the case of driven Rabi oscillations in weak external potentials. We conclude the applications by and example and apply this result to the case of a particle in a harmonic potential.

Lastly, we recap and conclude in Sec.~\ref{sec:conclusion} and comment on possible future applications for the results presented here.

\section{Decoupling Driven Rabi Models In External Fields}\label{sec:decouple}

Throughout this article we will employ the following notation: 
Operators on the system's total Hilbert space, consisting of the internal and external degrees of freedom, will be denoted with script letters (e.g. $\mathcal{H},\, \mathcal{S}$).
Operators on the external space will be denoted with a hat on top, for example the position operator $\hat{z}$ and the momentum operator $\hat{p}$.
We will often suppress the functional dependency of not explicitly time dependent objects like Heisenberg picture operators. 
Additionally, we will formally write sequences of operators containing an inverse as fractions where the denominator is always assumed to be ordered to the left hand side of the sequence. For example we have ${\mathcal{A}/\mathcal{B} = \mathcal{B}^{-1}\mathcal{A}}$ for two operators $\mathcal{A}$ and $\mathcal{B}$. With these preliminaries settled we turn to our task at hand, decoupling internal and external motion of a driven two-level system.

The Hamiltonian of a two-level system with internal states $\ket{e}$ and $\ket{g}$ with quantized motional degrees of freedom in presence of an external, classical laser field in paraxial and dipole approximation takes the form~\cite{Marzlin1996Feb,Lammerzahl1995Jul}
\begin{align}
    \mathcal{H}_0 = \sum\limits_{\mathclap{j\in\{\text{e},\text{g}\}}}\;\qty(\frac{\hat{p}^2}{2m} + V_j(\hat{z}))\otimes\dyad{j} -  E(\hat{z},t)\otimes\sum\limits_{\mathclap{\ell,j\in\{\text{e},\text{g}\}}} d_{\ell j}\ketbra{\ell}{j},
\end{align}
where $\hat{p}$ is the momentum operator and $\hat{z}$ the position operator of the two-level system along the propagation direction of the light field, $m$ the atomic mass, ${E(t,\hat{z})=E_0 \cos{(\omega_\text{L} t-k\hat{z} +\varphi_\text{L}(t))}}$ the magnitude of the electric field of the laser and the dipolse transition element ${d_{\ell j} = \bra{\ell}d\ket*{j}}$ of the atom.
The atom's internal energy ${\hbar\varepsilon_j}$ is accounted for by the state dependent potential ${V_j(\hat{z}) = Q(\hat{z}) + \hbar\varepsilon_j}$.
Note, that this form of interaction is generic, in the sense that a similar expression can be written down for the magnetic dipole interaction or any bilinear coupling between the two-level atom and external modes.

With the help of the rotating-wave approximation and by transforming intot he interaction picture with respect to the internal energy levels ${\hbar\varepsilon_j}$, this Hamiltonian can be approximated ~\cite{Fewell2005Sep} in a simpler form as~\cite{Marzlin1996Feb,Lammerzahl1995Jul}
\begin{align}
    \label{eq:hamiltonian}
        \mathcal{H}_0 = \begin{pmatrix}
        \frac{\hat{p}^2}{2m}+\hat{Q}
        &  \hbar \Omega~\E^{\I(\varphi(t)+ k\hat{z})}\\
         \hbar \Omega ~\E^{-\I(\varphi(t)+ k\hat{z})} & \frac{\hat{p}^2}{2m}+\hat{Q}
    \end{pmatrix},
\end{align}
with the phase function $\varphi(t)$, the external energy $\hat{p}^2/(2m)+\hat{Q}$ where the potential $\hat{Q}$ is assumed to be state- and time-independent. 
While state-independence up to a constant difference is mandatory for the following the requirement on time-independence can be lifted by using appropriate time-ordered exponentials in a generalized version of the derivation which follows below and does not change the end result.

Note, that any additional constants on the diagonal of Eq.~\eqref{eq:hamiltonian} are transformed into the phase term $\varphi(t)$ by the appropriate unitary transformation into the interaction picture.
The dipole operator matrix element defines the Rabi frequency ${\hbar \Omega = - E_0\bra{\ell}d\ket*{j}}/2$. Here $E_0$ is the amplitude of the electric field and $d$ is the dipole operator.
Equation~\eqref{eq:hamiltonian} has a variety of applications and appears in various systems, e.g. single Raman diffraction~\cite{Kleinert2015Dec,Moler1992Jan,Giese2015,Sanz2015Mar,Fewell2005Sep}.
The laser frequency $\omega_{\text{L}}$, the laser phase $\varphi_{\text{L}}(t)$ and the level splitting $\varepsilon_j$ are then collected in the time-dependent function
\begin{align}
    \varphi(t) = t\Omega\qty(\varepsilon_{\text{e}}-\varepsilon_{\text{g}} - \omega_{\text{L}}) - \varphi_{\text{L}}(t).
\end{align}
In this manner constant state dependent potentials, such as state dependent light shifts, can be removed in full analogy to the internal energy difference between the states.

In order to simplify the notation and explicitly highlight the relevant time-scales we introduce the dimensionless time ${\tau = t \Omega}$. 
Furthermore we introduce a dimensionless pair of momentum and position operators that define the characteristic (dimensionless) Doppler frequency operator and recoil length operator in relation to the Rabi frequency via \cite{Marzlin1996Feb}
\begin{equation}
 \hat{\nu}(\hat{p}) = \frac{k\hat{p}}{m\Omega}
 \quad \text{and} \quad
 \hat{\zeta}(\hat{z})=  \frac{k\hat{z}}{2\omega_{\text{r}}}\label{eq:canonical_operators}
\end{equation}
with the recoil frequency $\omega_{\text{r}} = \hbar k^2/(2m\Omega)$.
These dimensionless operators fulfill the canonical commutation relation ${\qty[\hat{\zeta},\hat{\nu}]=\I}$.

With these operators we define the displacement operator ${\hat{D} = \exp{\I\qty( 2\omega_{\text{r}}\hat{\zeta}+\phi(\tau))}}$ with the (laser) phase ${\phi(\tau) = \varphi(\tau / \Omega)}$, which naturally shifts the Doppler operator by $2\omega_{\text{r}}$.
Similarly, the kinetic energy term takes the form ${\hat{T} = \hat{p}^2/(2m \hbar \Omega) = \hat{\nu}^2/(4\omega_{\text{r}})}$ while the external potential becomes ${\hat{V} = \hat{Q}\qty(\hat{z})/(\hbar \Omega) = \hat{Q}\qty(2\omega_{\text{r}}\hat{\zeta}/(k))/(\hbar \Omega)}$.
Using these definitions the Hamiltonian, Eq.~\eqref{eq:hamiltonian}, takes the compact form
\begin{align}
    \frac{\mathcal{H}_0}{\hbar\Omega} = \begin{pmatrix}
        \hat{T}+\hat{V} & \hat{D}\\
       \hat{D}^{\dagger} & \hat{T}+\hat{V}
    \end{pmatrix} = \mathcal{H}.\label{eq:hamiltonian2}
\end{align}
In order to obtain the time evolution generated by this Hamiltonian we need to solve the Schrödinger equation
\begin{align}
    \I \frac{\dd}{\dd \tau} \mathcal{U} = \mathcal{H} \mathcal{U},\label{eq:initial_schroedinger} \quad \text{and} \quad
    \mathcal{U}(0) = \mathds{1}.
\end{align}
Our aim in this section is to decouple the Schrödinger equation with respect to the internal states by deriving a diagonalized equation.
To do so, we start by removing all diagonal elements of the Hamiltonian Eq.~\eqref{eq:hamiltonian2} via a unitary transformation, that is
\begin{align}
    \mathcal{S} &= \hat{S}\otimes \mathds{1}_2 =\exp{-\I \tau \qty(\hat{T}+\hat{V})}\otimes\mathds{1}_2\label{eq:interactionpicture}
\end{align}
where $\hat{S}$ is the part of the transformation acting purely on the external degrees of freedom.
This transformation leads to an interaction picture where the diagonal terms of the Hamiltonian are removed in their entirety.
While initially counter-intuitive, we will see shortly that this purely off-diagonal form will become handy, when we consider a second order equation emerging from this Schrödinger equation.

The resulting interaction picture Schrödinger equation and its Hamiltonian are
\begin{align}
    \I \frac{\dd}{\dd \tau} \mathcal{U}_{\mathcal{S}} = \mathcal{H}_{\mathcal{S}} \mathcal{U}_{\mathcal{S}}
    \;\; \text{with} \;\; 
    \mathcal{U}_{\mathcal{S}}(0) = \mathds{1}
    \;\; \text{and} \;\;
    \mathcal{H}_{\mathcal{S}} = \begin{pmatrix}
        0 & \hat{D}_{\text{S}} \\
        \hat{D}^{\dagger}_{\text{S}} & 0
    \end{pmatrix}.    
    \label{eq:initial_schroedinger}
\end{align}
This Hamiltonian $\mathcal{H}_ {\mathcal{S}}$ is governed by the operator ${\hat{D}_{\text{S}} = \hat{S}^{\dagger}\hat{D}\hat{S}}$ which is the interaction picture operator with respect to the external motion's evolution operator $\hat{S}=\exp{-\I \tau(\hat{T}+\hat{V})}$.
Within this interaction picture external operators evolve under $\hat{S}$. In particular they follow the Heisenberg equations of motion 
\begin{align}
    \frac{\dd}{\dd \tau} \hat{O}_{\text{S}} = \I\qty[\hat{H}_{\text{S}},\hat{O}_{\text{S}}]\label{eq:heisenberg_eom}
\end{align}
with respect to the external Hamiltonian $\hat{H}_{\text{S}} = \hat{T}+\hat{V}$.

As we have aimed for, the remaining Hamiltonian Eq.~\eqref{eq:initial_schroedinger} is now purely off-diagonal and governed solely by the unitary operator $\hat{D}_{\text{S}}$ (and its inverse $\hat{D}_{\text{S}}^{\dagger}$).

In the next step we exploit this property to decouple the equations for the internal degrees by defining the Schrödinger differential operator ${\mathcal{L} = \qty(\I \frac{\dd}{\dd \tau} - \mathcal{H}_{\mathcal{S}})}$ and applying $\mathcal{L}^{\dagger}$ to ${\mathcal{L}\,\mathcal{U}_{\mathcal{S}} = 0}$. This way we derive a second order equation for the time evolution operator.
The corresponding initial conditions are derived by applying the operator $\mathcal{L}$ once on the initial conditions of the original Schrödinger equation,  Eq.~\eqref{eq:initial_schroedinger}. 
This procedure yields the second order problem
\begin{align}
    & \mathcal{L}^{\dagger}\mathcal{L}\,\mathcal{U}_{\mathcal{S}} = \qty[\frac{\dd^2}{\dd \tau^2} + \mathcal{H}_{\mathcal{S}}^2 + \I \qty(\frac{\dd}{\dd \tau} \mathcal{H}_{\mathcal{S}})]\mathcal{U}_{\mathcal{S}} = 0 \quad \text{with}\label{eq:pre_inversion}\\
        &\mathcal{U}_{\mathcal{S}}\qty(0) = \mathds{1}
        \quad \text{and} \quad \dot{\mathcal{U}}_{\mathcal{S}}\qty(0) = -\I\begin{pmatrix}
        0 & \hat{D}_{\text{S}}(0) \\
        \hat{D}_{\text{S}}^\dagger(0) & 0
    \end{pmatrix}.\label{eq:oscillator_initial_values}
\end{align}

Our aim is now to convert the operator sequence ${\mathcal{L}^{\dagger}\mathcal{L}\,\mathcal{U}_{\mathcal{S}}}$ into a diagonal form.
This can be achieved by applying a formal trick.
Namely, we formally multiply the inverse of the Hamiltonian $\mathcal{H}_{\mathcal{S}}^{-1} $ onto the Schrödinger equation, Eq.~\eqref{eq:initial_schroedinger}, from the left such that the time evolution operator can be expressed formally as  ${ \mathcal{U}_{\mathcal{S}} = \I \mathcal{H}_{\mathcal{S}}^{-1} \frac{\dd}{\dd \tau} \mathcal{U}_{\mathcal{S}}}$.
By design of the unitary transformation from Eq.~\eqref{eq:interactionpicture} and the unitarity of the displacement we find the relation ${\mathcal{H}_{\mathcal{S}}^2=\mathds{1}}$ and consequently ${\mathcal{H_{\mathcal{S}}}^{-1} = \mathcal{H}_{\mathcal{S}}}$.
Note, that the inversion of the Hamiltonian is possible, if it is bounded from below as the energy can be shifted such that $\mathcal{H}_{\mathcal{S}}$ is positive definite. In our case at hand $\mathcal{H}_{\mathcal{S}}$ has eigenvalues that are $\pm 1$ independently of the potential that is assumed. Therefore,  an inversion is always possible.

Finally, by inserting the inverted Schrödinger equation into the last term of Eq.~\eqref{eq:pre_inversion} we arrive at 
\begin{align}
\mathcal{L}^{\dagger}\mathcal{L} \,\mathcal{U}_{\mathcal{S}} = \qty[\frac{\dd^2}{\dd \tau^2} + \mathds{1} - \qty(\frac{\dd}{\dd \tau} \mathcal{H}_{\mathcal{S}})\mathcal{H}_{\mathcal{S}}\frac{\dd}{\dd \tau}]\mathcal{U}_{\mathcal{S}} = 0\label{eq:post_inversion}
\end{align}
which is the promised second order equation.
The insertion of the inverse seems rather arbitrary and a bit ad-hoc at first but will ultimately lead to a diagonal form.

We are left with the task to evaluate the derivative of the Hamiltonian $\mathcal{H}_{\mathcal{S}}$ that is determined by unitary interaction picture displacements $\hat{D}_{\text{S}}$ and its adjoint.

By applying the usual Lie-algebraic formulas~\cite{Schur1891Jun,Rossmann2006Jun} or equivalent operator derivative rules~\cite{Wilcox1967} we obtain the derivative of the  interaction picture displacement as
\begin{align}
    \frac{\dd}{\dd \tau} \hat{D}_{\text{S}} = 2\I \hat{D}_{\text{S}}\hat{\delta}_{-}(\tau) = -2\I\hat{\delta}_{+}(\tau)\hat{D}_{\text{S}}.\label{eq:derivative_1}
\end{align}
For the details and intricacies of the operator derivative calculation we refer to Appendix \ref{appendix:section:derivative}. 
Furthermore, we have introduced the detuning operators $\hat{\delta}_{\pm}$ in the previous equation via
\begin{align}
  \hat{\delta}_{\pm}(\tau) = \frac{1}{2} \qty(\mp\frac{\dd\phi(\tau)}{\dd \tau} \mp  \hat{\nu}_{\text{S}} + \omega_{\text{r}})\label{eq:detuning}.
\end{align}
Here we also recall the definition of the dimensionless Doppler operator as defined in Eq.~\eqref{eq:canonical_operators}, $\hat{\nu}_{\text{S}} = \Omega^{-1} \frac{k\hat{p}_{\text{S}}}{m},$
as well as the (dimensionless) recoil frequency~\cite{Marzlin1996Feb,Lammerzahl1995Jul} via ${\omega_{\text{r}} = \Omega^{-1} \hbar k^2/(2m)}$.
We emphasize that the Doppler operator ${\hat{\nu}_{\text{S}} = \hat{S}^{\dagger}\hat{\nu}\hat{S}}$ is in the interaction picture  with respect to the evolution operator of the external motion $\hat{S}$ and therefore time-dependent.

In contrast to typical discussions of atomic diffraction based on resonant transitions and their slight off resonance due to a detuning~\cite{Marzlin1996Feb,Lammerzahl1995Jul,Moler1992Jan}, in our derivation the detuning and Doppler operators arise as natural consequences of the underlying operator structure without such assumptions and any notion of a diffraction process. 
Moreover, we observe that the  evolution of these operators is governed by the Heisenberg equations of motion, Eq.~\eqref{eq:heisenberg_eom}, with respect to the external generator of motion $\hat{H}_{S}$.

When we insert the derivative of the displacement, Eq.~\eqref{eq:derivative_1}, into the corresponding term of Eq.~\eqref{eq:post_inversion} we arrive at the identity
\begin{align}
    \qty(\frac{\dd}{\dd \tau}\mathcal{H}_{\mathcal{S}})\mathcal{H}_{\mathcal{S}} = - 2\I\begin{pmatrix}
     \hat{\delta}_{+}(\tau)&  0\\
    0 &  \hat{\delta}_{-}(\tau)
    \end{pmatrix}
\end{align}
which shows that the sequence ${\mathcal{L}^{\dagger}\mathcal{L}\,\mathcal{U}_{\mathcal{S}}}$ is diagonal with respect to the internal degrees of freedom as we have promised.

In general, the full time-evolution operator can be obtained from its matrix elements ${\hat{u}_{\ell j}=\bra{\ell}\mathcal{U}_{\mathcal{S}} \ket*{j}}$ with respect to the internal degree of freedom labeled by $\ell,j\in \{e,g\}$. However, usually these matrix elements are intricately coupled. In stark contrast, in our case the matrix elements solve separate and decoupled operator oscillation equations with respect to the external degrees of freedom only, that is
\begin{align}
    \qty[\frac{\dd^2}{\dd \tau^2}+ 2 \I  \hat{\delta}_{\pm}(\tau) \frac{\dd}{\dd \tau}+ 1]\hat{u}_{\ell j} = 0 \label{eq:oscillator}
\end{align}
where we use the index $+$ for ${\ell = \text{e}}$ and $-$ for ${\ell = \text{g}}$ in the individual elements.

This oscillator equation, Eq.~\eqref{eq:oscillator}, also accounting for the external dynamics is the key result of this article and will be applied to various examples in the course of Sec.~\ref{sec:applications}.
Furthermore, it is a damped harmonic oscillator where the damping terms are purely imaginary and their magnitude is determined by the hermitian detuning operators $\hat{\delta}_{\pm}(\tau)$ which in general are time-dependent.
This equation is not entirely unknown if the external motion is neglected and its solvability in this case is discussed extensively by Shore, e.g. in Ref.~\cite{Shore2011} in terms of the classical theory of special functions resulting from second order differential equations.

Note, arriving at this result depends crucially on inverting the Hamiltonian in the manipulations in-between Eq.~\eqref{eq:pre_inversion} and
 \eqref{eq:post_inversion}. Nevertheless, the only major restrictions we have imposed with respect to the starting Hamiltonian and during the derivation is that the operator part of the potential $\hat{Q}$ is common to both internal states and $\mathcal{H}_\mathcal{S}$ is invertible.

Moreover, our calculation holds for arbitrary phase functions $\phi(\tau)$ and thus the oscillator equation grants direct access to treating dynamic driving/laser phase variations or noise to which we return in Sec.~\ref{sec:laser_phase_noise}. 
Additionally, Eq.~\eqref{eq:oscillator} can serve an excellent basis for a perturbation expansion in cases where the operator valued detunings are small.
The interaction picture operators contained in the detuning usually do not pose an obstacle since the interaction picture momentum operator itself can be expanded perturbatively in various ways without solving the Heisenberg equation of motion exactly~\cite{Franson2002Apr,TeVrugt2019}.

Lastly, for a free particle and in a linear potential the oscillation equation becomes only momentum dependent and is hence an easily solvable equation, as shown explicitly in Sec.~\ref{subsec:no_potential} and Sec.~\ref{subsec:gravitation}.

\subsection{Equivalent Form of Oscillator Equations}\label{sec:equivalent-form}
Our result from Eq.~\eqref{eq:oscillator} can be cast into a form, where the first order derivative term $\dd/\dd \tau\,\hat{u}_{\ell j}$ is removed. In order to achieve this we take inspiration from the transformation theory of classical second order equations/Sturm-Liouville problems and integrating factors~\cite{Hassani}. However, in our case we need to add time/path-ordering since we are working with operator valued oscillator equations of the form
\begin{align}
\qty[\frac{\dd^2}{\dd \tau^2} + 2\I\hat{\delta}(\tau) \frac{\dd}{\dd \tau}+ 1]\hat{u} = 0
 \end{align}
for which it cannot be neglected. Note, that the previous equation is identical to Eq.~\eqref{eq:oscillator} if the $\pm$-indices are dropped.

As an \emph{integrating factor} we choose the unitary transformation defined by
\begin{align}
\hat{W} = \overline{\mathcal{T}}\exp\qty{\I\int\limits_{0}^{\tau} \!\! \dd s\, \hat{\delta}(s)}
\quad \text{and} \quad 
\hat  {u} = \hat{W} \hat{u}^{\prime}\label{eq:trafo}
\end{align}
where the overline signifies anti-time ordering leading to ${\D \hat{W}/ \D \tau = \I\hat{W} \hat{\delta}(\tau)}$.
As a consequence of this transformation we find the Sturm-Liouville normal form of our operator equation which becomes 
\begin{align}
\qty[\frac{\dd^2}{\dd \tau^2} +\hat{\beta}(\tau)] \hat{u}^{\prime} = 0  \label{eq:oscillator_noderivative}
\end{align}
for the operator $\hat{u}^{\prime}$ with the operator valued and time-dependent frequency
\begin{align}
\hat{\beta}(\tau) = 1 +\hat{\delta}^2(\tau)- \I\frac{\dd}{\dd \tau}\hat{\delta}(\tau).
\end{align}
Eq.~\eqref{eq:oscillator_noderivative} is free of first order time derivatives with respect to $\hat{u}^\prime$.

When we perform analogous transformations on all internal states of Eq.~\eqref{eq:oscillator} we arrive at an operator oscillation equation without the first derivative terms $\dd/\dd \tau\,\hat{u}_{\ell j}$, i.e.
\begin{align}
\qty[\frac{\dd^2}{\dd \tau^2} +\qty(1 + \hat{\delta}_{\pm}^2(\tau)-\I\qty(\frac{\dd}{\dd \tau}\hat{\delta}_{\pm}(\tau)))] \hat{u}^{\prime}_{\ell j} = 0.\label{eq:oscillator2}
\end{align}
This equation is a generalization to arbitrary potentials of the approach first put forward by Marzlin in Ref.~\cite{Marzlin1996Feb} in the case of a linear potential.
We observe that the detuning $\hat{\delta}_{\pm}$ emerges quadratically in combination with its first derivative in contrast to Eq.~\eqref{eq:oscillator} where it only appears linearly and its derivative is absent.

While fundamentally interesting, this result, as well as perturbation theory based upon it, is rather cumbersome since the reference oscillator now explicitly depends on time.
In contrast, in our original approach, the reference oscillator in Eq.~\eqref{eq:oscillator} is time-independent and has unit frequency.
Moreover, we do not need to compute the square of the detuning operator and its derivative.
For example, in case of detuning operators that are polynomial functions of $\hat{x}$ and $\hat{p}$ the squaring will lead to much more complicated operator dependencies.
Another formidable problem is posed by finding an explicit expression for the time-ordered exponential of the detuning in the transformation, Eq.~\eqref{eq:trafo}, which is needed to link this formulation to the original problem.
These reasons suggest that a generalization of the approach in Ref.~\cite{Marzlin1996Feb} is less suitable for a perturbative analytical treatment in arbitrary potentials.

In contrast, in our oscillator equation, Eq.~\eqref{eq:oscillator}, the detuning appears linearly and only the interaction picture Doppler operator needs to be calculated.
This allows for a straight forward implementation of perturbation theory within the oscillation equations as we explicitly show in Sec.~\ref{subsec:numerics}.

Before we apply Eq.~\eqref{eq:oscillator} to different potentials in Sec.~\ref{sec:applications} it is useful to analyze its basic properties and implications. More specifically, it will be helpful to obtain a basic understanding of the detuning operators, their action and how they can be reconnected to the intuitive idea of scalar detunings since they are the governing elements of the oscillator equations.
Therefore we will investigate the instantaneous eigenstates of the detuning operators and analyze the case where in a given setup the explicit time-dependence can be removed completely or approximately from the oscillation Eq.~\eqref{eq:oscillator}.

\subsection{Generalized Detuning Eigenspace and Algebra}\label{sec:resonance-condition-emerges}
First, we recall from Eq.~\eqref{eq:detuning} the detuning operator
\begin{equation}
\hat{\delta}_{\pm}(\tau) = \frac{1}{2} \qty(\mp\frac{\dd\phi(\tau)}{\dd \tau} \mp  \hat{\nu}_{\text{S}} + \omega_{\text{r}})\label{eq:detuning2}
\end{equation}
with the interaction picture operator $\hat{\nu}_{\text{S}} = \Omega^{-1}k \hat{p}_\text{S}/m$ and the phase function $\phi(\tau)$.
In typical treatments of atomic diffraction, as put forward for example in Refs.~\cite{Moler1992Jan,Giese2015}, the resonance condition is typically formulated via energy momentum conservation in discretized momentum space between the initial and final atomic state.

In our case we have to consider both detuning operators and their instantaneous eigenspaces. In order to treat both simultaneously we solve the generalized eigenvalue problem of the second kind~\cite{Golub2013} defined by
\begin{align}
    \begin{pmatrix}
        \hat{\delta}_+&0\\
        0& \hat{\delta}_-
    \end{pmatrix}\begin{pmatrix}
        \ket*{\nu_+}\\
        \ket*{\nu_-}
    \end{pmatrix} = \delta \sigma_{\text{z}} \begin{pmatrix}
        \ket*{\nu_+}\\
        \ket*{\nu_-}
    \end{pmatrix}.\label{eq:generalized_eigenvalue}
\end{align}
The solution of this generalized eigenvalue problem yields (simultaneous) eigenvectors not only for the detuning operator but at the same time for the internal spin operator $\sigma_{\text{z}}$.

Physically, the appearance of the operator $\sigma_{\text{z}}$ can be traced back to the fact that before and after the interaction has been switched on and off the system is in a well defined spin as well as motional state. Alternatively, and guided by the math, we have simply chosen the eigenbasis of $\sigma_{\text{z}}$ as our matrix representation for the two-level system.

Under the assumption that $\ket*{\nu_{\pm}}$ are eigenfunctions of $\hat{\nu}_{\text{S}}$ with the eigenvalues $\nu_{\pm}$, we find from Eq.~\eqref{eq:detuning2} and Eq.~\eqref{eq:generalized_eigenvalue} the relation
\begin{align}
\nu_+ = \nu_- + 2 \omega_{\text{r}}
\end{align}
or equivalently in momentum space ${p_+ = p_- + \hbar k}$,
which shows, that the momentum change from $p_-$ to $p_-+\hbar k$ is connected with an internal state transition. Moreover, the limit $\delta \to 0$ and the eigenvalue $\delta =0$ determines what is usually called the resonant Doppler frequency $\nu_0$ via the (resonance) condition
\begin{align}
\begin{pmatrix}
        \hat{\delta}_+&0\\
        0& \hat{\delta}_-
    \end{pmatrix}\begin{pmatrix}
        \ket*{\nu_0+2\omega_{\text{r}}}\\
        \ket*{\nu_0}
    \end{pmatrix} = \Vect{0}\label{eq:resonance_Doppler}
\end{align}
or equivalently in terms of the momentum operator and its eigenfunctions the resonant momentum $p_0$ via
\begin{align}
\begin{pmatrix}
        \hat{\delta}_+&0\\
        0& \hat{\delta}_-
    \end{pmatrix}\begin{pmatrix}
        \ket*{p_0+\hbar k}\\
        \ket*{p_0}
    \end{pmatrix} = \Vect{0}.\label{eq:resonance}
\end{align}
We emphasize that the resonance condition is fulfilled for specific pairs of eigenvectors residing in the individual kernels of the respective detuning operators which can correspond to multiple or even infinitely many values.

\subsection{Modifiying the Detuning Operator and Chirping}
\label{sec:chirping-and-phase-noise}
By adjusting the time-dependence $\phi(\tau)$ of the phase of the driving field we can modify and influence the detuning operator. This procedure~\cite{Giese2015} is widely known as chirping the resonance. More specifically, if we force the average detuning to be constant in time via the condition
\begin{align}
    \frac{\dd}{\dd \tau}
    \big\langle \hat{\delta}_\pm \big\rangle=
    \frac{\dd}{\dd \tau}\big\langle\mp\frac{\dd \phi(\tau)}{\dd \tau} \mp  \hat{\nu}_{\text{S}}  + \omega_{\text{r}}\big\rangle = 0
\end{align}
we arrive at the chirping condition
\begin{align}
    \label{eq:chirp-phase-1d}
    &\frac{\dd^2 }{\dd  \tau^2}\phi(\tau) =  -\frac{\dd}{\dd \tau}\langle\hat{\nu}_{\text{S}}\rangle = \frac{k}{m\Omega^2}\Big\langle\frac{\partial V(x)}{\partial x}\Big\vert_{x = \hat{x}_{\text{S}}(\tau)}\Big\rangle
\end{align}
where $V(x)$ is the potential occuring in the external Hamiltonian $\hat{H}_{\text{S}}$.
More generally, an analogous consideration to our calculation in three dimensions leads to the general condition
\begin{align}
    \label{eq:chirp-phase-3d}
    &\ddot{\varphi}(t) = \frac{ \Vec{k}}{m}\cdot\Big\langle \nabla_{\vec{x}}V(\vec{x})\Big\vert_{\vec{x} = \hat{\vec{x}}_{\text{S}}(t)} \Big\rangle    
\end{align}
after using ${\tau=\Omega t}$ and recalling ${\varphi(t)=\varphi\big(\tau(t)/\Omega\big)\equiv \phi(\tau)}$. This condition needs to be paired with appropriate initial conditions, e.g. ${\dot{\varphi}(0)=\Delta\omega}$ and ${\varphi(0)=\varphi_0}$.
In atom interferometry experiments the freedom of choosing the phase $\varphi$ is often used~\cite{Peters2001Feb} to compensate for the leading order (Ehrenfest) contributions of the atomic motion~\cite{Kleinert2015Dec} entering from the detuning operator. More specifically, and as we show in Sec.~\ref{subsec:gravitation}, in case of a linear potential ${\hat{V}=\Vect{a}\Vect{r}}$ the effect of the potential can be removed completely by choosing ${\varphi(t)=\varphi_0+\Delta\omega t+\Vect{k} \Vect{a} t^2/(2m)}$. 
In all other cases one needs to content oneself with an approximate removal of the average detuning.

\subsection{Laser Phase Noise}\label{sec:laser_phase_noise}
Before moving on to applications we briefly discuss the impact of phase noise, namely the case where the driving fields phase $\varphi(t)$ is replaced via $\varphi(t)\mapsto \varphi(t)+\delta\varphi_t$ where $\delta \varphi_t$ is a stochastic process with appropriate statistics, e.g. a Wiener process with $\langle \frac{\dd}{\dd t}\delta \varphi_t\rangle_t =0$ and $\langle \frac{\dd}{\dd t}\delta\varphi_t \frac{\dd}{\dd t^\prime}\delta\varphi_{t^\prime}\rangle_t = s \delta(t-t^\prime)$, where $\langle \bullet \rangle_t$ denotes the time-average of a generic function represented by the placeholder $\bullet$.
Hence it is straight forward to include laser phase noise in Eq.~\eqref{eq:chirp-phase-1d} and Eq.\eqref{eq:chirp-phase-3d} which leads to a term proportional to $ \delta \ddot{\varphi}_t$ and thus a stochastic violation of the chirping condition.
Note that the second derivative of the stochastic process is to be understood in a distributional sense~\cite{Zemanian1965} and only makes sense for processes where it is well defined.
However, even in this more complicated case Eq.~\eqref{eq:chirp-phase-1d} and Eq.~\eqref{eq:chirp-phase-3d} can be used to determine an on average valid chirp $\phi(t)$ by stochastically integrating the system with its Green's function as presented for the noisy non-linear classical oscillator in Ref.~\cite{Gitterman2005Nov} or by using other techniques to integrate this stochastic differential equation~\cite{Evans2013}.

\section{Applications}\label{sec:applications}
In the present section we will apply our oscillator decomposition, Eq.~\eqref{eq:oscillator}, to three different examples. In particular, we will begin with the case of a free particle, that is the case of vanishing external potential $\hat{Q}\equiv 0$ in Sec. \ref{subsec:no_potential}. This example is directly followed by the linear potential ${\hat{Q}=a\hat{z}}$ in Sec.~\ref{subsec:gravitation}.
Finally, in Sec.~\ref{subsec:numerics}, we will treat a weak quadratic potential $\hat{Q}=\epsilon \alpha \hat{z}^2$ with $\alpha$ being a dimensionful constant which could correspond to, e.g. a gravity gradient.
In case of the latter we will derive a general scheme on how to treat arbitrary potentials semi-analytically in a perturbative scheme. 

\subsection{Analytic Solution for Atomic Diffraction without Potentials}\label{subsec:no_potential}
In the absence of any external potentials we have ${\hat{Q} = 0 }$ and as a consequence the Heisenberg momentum operator is identical to the Schrödinger operator. Thus we have ${\hat{p}_{\text{S}} = \hat{p}}$.
For the rate of change of the phase $\phi(\tau)$ we choose a constant ${\dd \phi(\tau)/\dd  \tau = \Delta\omega}$.

Therefore, the detuning operator, Eq.~\eqref{eq:detuning}, reduces to the time-independent operator ${\hat{\delta}_{\pm} = \mp \qty(\Delta\omega +  \hat{\nu} \mp \omega_{\text{r}})/2}$ and we arrive at the oscillation equation
\begin{align}
\qty(\frac{\dd^2}{\dd \tau^2} +2\I\hat{\delta}_{\pm}\frac{\dd}{\dd \tau} + 1 )\hat{u}_{\ell j} = 0.\label{eq:no_potential}
\end{align}
Since Eq.~\eqref{eq:no_potential} is governed by a single time-independent hermitian operator, namely the momentum operator $\hat{p}$, we can solve it by expanding it in the momentum eigenbasis. As a consequence we only need to solve a classical equation for every eigenstate. Alternatively, and even more simple we note that the operator nature can be disregarded completely here since the momentum operator commutes with itself at all times.

Regardless of the way chosen, the solution of Eq.~\eqref{eq:no_potential} is identical to the well-known expression for a damped classical oscillator
\begin{align}
    \hat{u}_{\ell j} = \E^{\I \hat{\delta}_{\pm} \tau} \qty(\cos(\hat{\mu}_{\pm} \tau)\hat{C}_1+ \sin(\hat{\mu}_{\pm} \tau)\hat{C}_2)
\end{align}
with the effective Rabi frequency operator $\hat{\mu}_\pm$ defined by
\begin{align}
    \hat{\mu}_{\pm} = \sqrt{1+\hat{\delta}_{\pm}^2}
    =\sqrt{1+\big(\Delta \omega +\hat{\nu}\mp \omega_\text{r}\big)^2/4}.
\end{align}
Moreover, with the help of the initial conditions from~Eq.~\eqref{eq:oscillator_initial_values} we can directly obtain the operators $\hat{C}_1$ and $\hat{C}_2$ and thus arrive at the expression
\begin{widetext}
    \begin{align}
\mathcal{U}_{\mathcal{S}} = \begin{pmatrix}
    \E^{-\I \hat{\delta}_+ \tau} & 0 \\
    0 & \E^{-\I \hat{\delta}_-\tau}
\end{pmatrix}
\begin{pmatrix}
    \cos(\hat{\mu}_+ \tau) + \I \sin(\hat{\mu}_+ \tau) \frac{\hat{\delta}_+}{\hat{\mu}_+} & - \I \sin(\hat{\mu}_+ \tau) \frac{\Kick_{\text{S}}(0)}{\hat{\mu}_+} \\
     - \I \sin(\hat{\mu}_- \tau) \frac{\Kickdag_{\text{S}}(0)}{\hat{\mu}_-} & \cos(\hat{\mu}_- \tau) + \I \sin(\hat{\mu}_- \tau) \frac{\hat{\delta}_-}{\hat{\mu}_-}\label{eq:rabi}
\end{pmatrix}
\end{align}
\end{widetext}
for the time evolution operator $\mathcal{U}_{\mathcal{S}}$ in the interaction picture.
We emphasize again that the expression $\hat{D}/\hat{\mu}$ is a formal notation and synonymous with the sequence $\hat{\mu}^{-1}\hat{D}$.
After inspecting Eq.~\eqref{eq:rabi}, we note the stark similarity to a detuned Rabi oscillation. However, now with the detuning depending on quantized external degrees of freedom. The momentum-dependent matrix elements resulting from Eq.~\eqref{eq:rabi} are plotted in Fig. \ref{fig:phasenosie_0}.
\begin{figure}
    \centering
    \includegraphics[width = \columnwidth]{figures/densityplot_no_potential_low.png}
    \caption{Solution of the oscillator equation, Eq.~\eqref{eq:no_potential}, with no potential. Density plots of the evolution amplitudes $\hat{u}_{\ell j}\hat{u}_{\ell j}^{\dagger}$ and phase of the elements $u_{\text{ee}}$ in {\sffamily (a)}, $u_{\text{eg}}$ in {\sffamily (b)}, $u_{\text{ge}}$ in {\sffamily (c)} and $u_{\text{gg}}$ in {\sffamily (d)} as a function of momentum and pulse time. The left column side shows only the absolute value of the evolution operator matrix element $\hat{u}_{\ell j}(\hat{\nu},\tau)$ on the eigenspace of $\hat{\nu}$. The right column displays the respective argument of the matrix element. We have mapped the transparency to the absolute value of the element $u$. Hence, bright white areas in the phase diagramms correspond to vanishing amplitude. The dashed lines mark the $\pi/2$- and $\pi$-pulses, that are of particular interest in most applications. We chose ${\dd\phi(\tau) /\dd \tau  = \Delta\omega =  \omega_{\text{r}} = 1}$ in our examples. For momenta further from the resonant momentum pair, that is ${\nu=0}$ for the excited state and ${\nu=-2\omega_{\text{r}}}$ for the ground state, we observe faster oscillations as the oscillation frequency is determined by ${\mu(\nu) = \sqrt{1+\delta^2(\nu)}}$.}
    \label{fig:phasenosie_0}
\end{figure}

Note, that up to now we are still in an interaction picture with respect to the external motion. In order to obtain the evolution in the laboratory frame evolution we reverse the interaction pictures via\footnote{In general, the transformation back into the lab frame, where the evolution starts at $\tau_0$ is given by $\mathcal{U}(\tau) = \mathcal{R}(\tau)\mathcal{S}(\tau)\mathcal{U}_{\mathcal{S}}(\tau)\mathcal{S}^{\dagger}(\tau_0)\mathcal{R}^{\dagger}(\tau_0)$.}
\begin{align}
\mathcal{U}(\tau)&= \mathcal{R}(\tau)\mathcal{S}(\tau)\mathcal{U}_{\mathcal{S}}(\tau)\label{eq:back_to_lab}
\end{align}

with the diagonal operators ${\mathcal{S}(\tau)=\E^{-\I\hat{\nu}^2\tau/(4\omega_r)}\otimes \mathds{1}_2}$ and ${\mathcal{R}(\tau)=\text{diag}(\E^{-\I\varepsilon_e \tau},\E^{-\I\varepsilon_g \tau})}$ accounting for the free external and internal evolution, respectively.
In case of the free particle each of these matrices can be evaluated efficiently in momentum representation for arbitrary pulse duration. The result in Eq.~\eqref{eq:rabi} is the representation free evolution operator version of the expressions for single Raman diffraction, previously obtained in Refs.~\cite{Moler1992Jan,Giese2015} in terms of momentum space probability amplitudes. In contrast to these treatments the previous expression can be applied to arbitrary states.

\subsection{Analytic Solution for Atomic Diffraction in Linear Potentials}\label{subsec:gravitation}
Similarly to the case of vanishing potential, in case of a linear potential with ${\hat{V} = a \hat{z}/(\hbar \Omega)}$ the Heisenberg  operator for momentum becomes very simple and obeys ${\hat{p}_{\text{S}}(\tau) = \hat{p} - a/\Omega\,\,\tau}$ or equivalently in dimensionless form ${\hat{\nu}_{\text{S}}(\tau)=\hat{\nu}-2 \kappa \tau}$ with ${\kappa=ka/(2m\Omega)}$.
When we furthermore define the abbreviation ${\kappa_{\pm} = \pm \kappa}$ we arrive at the oscillator equations
\begin{align}
    \qty[\frac{\dd^2}{\dd \tau^2} +2\I\qty(\hat{\delta}_{\pm}+\kappa_{\pm}\tau)\frac{\dd}{\dd \tau}+1]\hat{u}_{\ell j} = 0, \label{eq:oscillator_gravity}
\end{align}
where ${\hat{\delta}_{\pm} =\mp \qty(\Delta\omega +  \hat{\nu}\mp\omega_{\text{r}})/2}$ is identical to the case from Eq.~\eqref{eq:no_potential} where no potential was present.
In case of a linear potential we have ${[\hat{\delta}(\tau),\hat{\delta}(\tau^\prime)]=0}$ and thus the relevant transformation is $\hat{W}_\pm=\exp{\mp\I\qty(\Delta \omega \tau + \hat{\nu}\tau + \kappa \tau^2\mp \omega_r \tau)/2}$.

The solutions of Eq.~\eqref{eq:oscillator_gravity} are given by a combination of the confluent hypergeometric functions and gamma functions~\cite{Polyanin2002Oct,Marzlin1996Feb,Lammerzahl1995Jul}.
After inserting the initial conditions, Eq.~\eqref{eq:oscillator_initial_values}, we arrive at
\begin{align}
    \hat{u}_{\ell j} = -\frac{2\kappa_{\pm}}{\I2\kappa_{\pm}-1}
    \exp{-2\I\hat{\delta}_{\pm}\tau-\I\kappa_{\pm}\tau^2}
    \frac{A\qty(\kappa_{\pm},\hat{\delta}_{\pm},\tau)}
    {B\qty(\kappa_{\pm},\hat{\delta}_{\pm},\tau)}\hat{D}^{\mp}
\end{align}
for $\ell\neq j$ and
\begin{align}
    \hat{u}_{\ell \ell} =  \exp{-2\I\hat{\delta}_{\pm}\tau-\I\kappa_{\pm}\tau^2} \frac{    C\qty(\kappa_{\pm},\hat{\delta}_{\pm},\tau)
    }
    { 
    (2\I\kappa_{\pm}-1)G\qty(\kappa_{\pm},\hat{\delta}_{\pm},\tau)
    }
\end{align}
for the matrix elements of the residual time-evolution operator.

The functions $A$, $B$, $C$ and $G$ are defined in Appendix~\ref{appendix:subsec:linear_potential} in terms of special functions.
The plus and minus signs are defined via $+$ if ${\ell 
 = \text{e}}$ and $-$ if ${\ell =\text{g}}$.
In Fig.~\ref{fig:analytic2} we exhibit density plots of amplitude and phase of the resulting matrix-elements of the time-evolution operator.

\begin{figure*}[t]
    \centering
    \includegraphics[width = \textwidth]{figures/densityplot_linear_low.png}
    \caption{Solution of Eq.~\eqref{eq:oscillator_gravity} for the element $u_{\text{ee}}$ in (a) and $u_{\text{eg}}$ in (b). We show the complex eigenvalue of the element $u$ with respect to the Doppler operator $\hat{\nu}$ and the time $\tau$. We chose $\kappa = -1$, ${\omega_{\text{r}}=0.1}$ and ${\Delta\omega = 0}$. 
    The resonance is shifted by the linear potential which changes the detuning linearly in time. Therefore, the resonant momentum is changed over time which leads to the linear downwards shift of the resonant momentum. 
    }
    \label{fig:analytic2}
\end{figure*}

Note, that the effect of the linear potential on the Rabi oscillation can be compensated right out of the gate by setting $\Delta \omega \mapsto -(\nu_0+\omega_\text{r}) - 2 \kappa \tau $, which corresponds to chirping the phase via setting $\phi(\tau)=\phi(0)-(\nu_0+\omega_\text{r}) \tau -\kappa \tau^2$ since $\D \phi/\D\tau = \Delta \omega$ where $\nu_0$ and $\nu_0+2\omega_\text{r}$ or equivalently $p_0$ and $p_0+\hbar k$ are the resonantly linked momenta as shown by Eq.~\eqref{eq:resonance_Doppler} and Eq.~\eqref{eq:resonance}, respectively. This laser phase adjustment corresponds to what is typically done in experiments when chirping is applied to stay on resonance~\cite{Chiow2011Sep}.
The elements time evolution operator $\mathcal{U}_{\mathcal{S}}$ is shown in Fig.~\ref{fig:analytic2}.

Lastly, the transformation back into the laboratory frame is performed analogously to Eq.~\eqref{eq:back_to_lab} except for the slightly different external operator ${\mathcal{S}(\tau) = \E^{-\I\tau(\hat{\nu}^2/(4\omega_r)+2 \kappa \hat{\zeta})}\otimes \mathds{1}_2}$.
Here we make use of the dimensionless position operator ${\hat{\zeta} = \hat{z} m \Omega/(\hbar k)}$ as introduced in Sec.~\ref{sec:decouple}.

The exponential in $\mathcal{S}$ can be decomposed into individual exponentials by applying the Baker-Campbell-Hausdorff formula yielding
${\mathcal{S}(\tau) = \E^{-\I  \kappa^2\tau^3/(12\omega_{\text{r}})}\E^{-\I 2 \kappa \hat{\zeta}\tau}\E^{-\I\tau\qty(\hat{\nu}-\kappa\tau )^2/(4\omega_{\text{r}})}\otimes \mathds{1}_2}$.
The term that contains the position operator $\hat{\zeta}$ can be identified as displacements that shift to the classical momentum, which the particle acquires due to the external potential acting for the pulse time $\tau$.
This effect cannot be compensated by chirping but is common to both states and hence in a globally linear potential does not lead to differential phases. 

Finally, we emphasize that Eq.~\eqref{eq:oscillator_gravity} can be mapped into the result derived by Marzlin et al. in Ref.~\cite{Marzlin1996Feb} by a unitary transformation as sketched in Sec.~\ref{sec:equivalent-form} and is thus equivalent to it. In light of this equivalence we also note that our result agrees perfectly with the one found by Lämmerzahl et al. in Ref.~\cite{Lammerzahl1995Jul}. However, again our result is formulated in a representation free description making it much more widely applicable.

\subsection{Efficient Numerical Implementation of Atomic diffraction in Potentials}\label{subsec:numerics}
In general potentials the interaction picture operator $\hat{\nu}_{\text{S}}$ is time-dependent and might mix the Schrödinger operators $\hat{\zeta}$ and $\hat{\nu}$ arbitrarily. As a consequence an analytic solution containing only one of the canonical operators, leading to the simplifications seen in the previous sections, is often not forthcoming. 
However, in case of a weak perturbing potential of the form $\hat{V} = \epsilon U(\hat{\zeta})$ to any of the previous cases, a perturbative scheme can be constructed. The development of this approach for an initial wave function and its exemplary application to a weak harmonic potential is the subject of this section.
Here we choose the relatively simple solution of Rabi oscillations without potentials derived in Sec.~\ref{subsec:no_potential} as a perturbative basis.
This is guided by the fact, that numerical evaluation of the expressions for the free particle solution is cheaper than the complicated special function expressions in case of the linear potential solution.
Furthermore, instead of using the oscillation equation for the time evolution operator we will use the corresponding expressions for the quantum states instead, since this is more accessible to numerics and interpretation.

The analogue oscillation equation for the quantum states follows directly from considering ${\mathcal{L}^{\dagger}\mathcal{L}\ket*{\psi} = 0}$ in full analogy to Sec.~\ref{sec:decouple} and leads to the same oscillation equation as in Eq.~\eqref{eq:oscillator}. However, the matrix elements of the time evolution operator are replaced by the states ${\ket*{\psi_\pm}=\bra{\pm}\ket*{\psi}}$ with respect to the eigenbasis $\ket*{\pm}$ of $\sigma_z$ corresponding to spin-up $(+)$ and spin-down $(-)$.

We set up the perturbative scheme by expanding the 
 interaction picture detuning operator and the quantum state in terms of a small parameter $\epsilon$ via
\begin{align}
    \hat{\delta}_{\pm,\text{S}} = \hat{\delta}_{\pm} \pm  \sum\limits_{k=1}^{\infty} \epsilon^k\hat{\delta}^{(k)}(\hat{\zeta},\hat{\nu})
    \quad \text{and} \quad 
    \ket*{\psi_{\pm}} = \sum\limits_{k=0}^{\infty} \epsilon^k \ket*{\psi_{\pm}^{(k)}}.
\end{align}
After collecting terms of equal powers in $\epsilon$ we arrive at the infinite set of equations
\begin{align}
    \begin{split}
    &\qty[\frac{\dd^2}{\dd \tau^2}+1+2\I\hat{\delta}_{\pm}\frac{\dd}{\dd \tau}]\ket*{\psi_{\pm}^{(0)}} = 0\\
    &\qty[\frac{\dd^2}{\dd \tau^2}+1+2\I\hat{\delta}_{\pm}\frac{\dd}{\dd \tau}]\ket*{\psi_{\pm}^{(k)}} = \mp 2 \I \sum\limits_{l=0}^{k-1} \hat{\delta}
    ^{(k-l)}\frac{\dd}{\dd \tau}\ket*{\psi_{\pm}^{(l)}}.
    \end{split}
    \label{eq:perturbation_theory_1}
\end{align}
Note, that we can solve the lowest order equation in Eq.~\eqref{eq:perturbation_theory_1} as shown in Sec.~\ref{subsec:no_potential}.
In order to obtain a solution of order $k$ in momentum space we solve Eq.~\eqref{eq:perturbation_theory_1} iteratively by using the classical retarded Green's function
\begin{align}
    G_{\pm}(\tau,\tau^{\prime},\nu) =  \frac{\theta(\tau-\tau^{\prime})}{2\mu_{\pm}(\nu)}\E^{-\I \delta_{\pm}\qty(\tau-\tau^{\prime})}\sin{\big(\mu_{\pm}(\nu)\qty(\tau-\tau^{\prime})\big)}.
\end{align}
The wave function of any higher order is obtained by performing the integration
\begin{align}
    \psi_{\pm}^{(k)}(\nu,\tau) = \mp 2\I \sum\limits_{l = 0}^{k-1} \!\int\limits_{0}^{\tau}\!\!\dd \tau^{\prime} \, G_{\pm}(\tau,\tau^{\prime},\nu) \bra*{\nu}\hat{\delta}
    ^{(k-l)}\frac{\dd}{\dd \tau^{\prime}}\ket*{\psi_{\pm}^{(l)}(\tau^{\prime})}.\label{eq:perturbation_solution}
\end{align}
In order to apply the operators $\hat{\delta}^{(k-l)}(\hat{\zeta},\hat{\nu})$ to $\psi_{\pm}^{(l)}(\nu,s)$ one can Fourier transform back and forth between the dual space to apply terms that contain the position operator $\hat{\zeta}$ or momentum operator $\hat{\nu}$ efficiently.
The perturbative scheme, Eq.~\eqref{eq:perturbation_theory_1}, and its solution Eq.~\eqref{eq:perturbation_solution} are the second central results of this article. 

 \subsubsection{Quadratic Potential}\label{subsec:quadratic_potential}
 In this section we apply the perturbative formalism to a quadratic potential, i.e. ${\hat{V} = \epsilon \hat{\zeta}^2}$.
 By solving the Heisenberg equations of motion we find
 \begin{align}
    \hat{\nu}_{\text{S}} & =\cos{\left(\varphi \tau\right)}\hat{\nu} - \frac{\epsilon}{\varphi}\sin{\left(\varphi \tau\right)}\hat{\zeta},\quad \varphi = \sqrt{\frac{\epsilon}{\omega_{\text{r}}}}
\end{align}
and therefore, when expanded in orders of $\epsilon$ we obtain
\begin{align}
\begin{split}
    \hat{\nu}_{\text{S}} 
    =\hat{\nu} - 2\sum\limits_{k=1}^{\infty} \epsilon^{k} \hat{\delta}^{(\ell)}
 \end{split}
 \end{align}
 with the definition $\hat{\delta}^{(\ell)} = g_\ell(\tau)\hat{\nu}/2+h_\ell(\tau)\hat{\zeta}/2$ as well as the abbreviations
\begin{align}
    \begin{split}
    g_k(\tau) &= -\frac{(-1)^k}{(2k)!}\frac{\tau^{2k}}{\omega_{\text{r}}^k}, \\
    h_k(\tau) &= 2\frac{(-1)^{k-1}}{(2k-1)!}\frac{\tau^{2k-1}}{\omega_{\text{r}}^{k-1}}
    \end{split}
\end{align}
for the series coefficients.
 
We solve Eq.~\eqref{eq:perturbation_theory_1} order by order numerically via the evaluation of Eq.~\eqref{eq:perturbation_solution}.
Afterwards we transform back to the laboratory frame by applying ${\mathcal{S}(\tau) = \E^{-\I\tau(\hat{\nu}^2/(4\omega_r)+\epsilon\hat{\zeta}^2)}\otimes \mathds{1}_2}$. This can be done in various ways: One is decomposing the wave functions into the eigenfunctions of the quantum harmonic oscillator. Alternatives are the application of the time evolution operator for the harmonic oscillator directly in momentum representation or numerical split stepping methods.
The probability and phase density results for the propagation of a Gaussian initial distribution with different potential strengths for $\alpha$-pulses with $\alpha \in \{0,2\pi\}$ are shown in Fig.~\ref{fig:quadratic} and showcase the combined action of Rabi oscillation between internal states and the harmonic oscillator potentials.

 \begin{figure*}[b]
     \centering
     \includegraphics[width = 0.49\textwidth]{figures/densityplot_quadratic_1_low.png}
     \includegraphics[width = 0.49\textwidth]{figures/densityplot_quadratic_2_low.png}
     \caption{Left panel {\sffamily (a)} and {\sffamily (b)}:  Full time evolution of a gaussian wave position wave packet with respect to Eq.~\eqref{eq:hamiltonian} in a quadratic potential. After solving the perturbative set Eq.~\eqref{eq:perturbation_theory_1} we applied the evolution of the external potential via decomposing the wave function into the eigenfunctions of the harmonic oscillator and adding the corresponding phase. The semi-analytical solution for the element $u_{\text{ee}}$ in {\sffamily (a)} and $u_{\text{eg}}$ in {\sffamily (b)} in presence of a quadratic potential up to the perturbative order ${k=10}$ via Eq.~\eqref{eq:perturbation_theory_1} with ${\epsilon= 10^{-3}}$, ${\omega_{\text{r}} = 0.5}$, ${\Delta\omega = -\omega_{\text{r}}}$ for a Gaussian initial wave packet centered at $\nu = 0 $ with a with of ${\sigma = 0.1}$. 
     Right panel {\sffamily (c)} and {\sffamily (d)}:~The solution of Eq.~\eqref{eq:hamiltonian} for another set of parameters is shown for $u_{\text{ee}}$ in {\sffamily (c)} and $u_{\text{eg}}$ in {\sffamily (d)}. We solved up to order ${k=10}$ via Eq.~\eqref{eq:perturbation_theory_1} with ${\epsilon= 6\cdot 10^{-2}}$, ${\omega_{\text{r}} = 0.25}$, ${\Delta\omega = -\omega_{\text{r}}}$ and  a Gaussian initial wave packet centered at ${\nu = 0 }$ with a standard deviation of ${\sigma = 0.25}$.
     The asymmetry in the diffraction process, especially the ongoing population in the ground state for momenta that are larger than zero can be explained by the combination of resonant diffraction and external motion. In {\sffamily (b)} the transition from ${\nu = 0}$ to ${\nu = 1}$ is resonant. Therefore the slightly more positive momenta gain a higher momentum during the diffraction process. 
     Since the wave packet is in an external quadratic potential the external motion reduces the momentum, such that the momenta that are larger than ${\nu = 1}$ move closer to the resonant transition and have an overall lower detuning than momenta that start slightly lower than ${\nu = 0}$ at ${\tau = 0}$. Those parts of the wave function fall even further from resonance when their momentum is reduced in the external potential while being diffracted, such that they move away from resonance. That is why smaller momenta than ${\nu = 0}$ are more off-resonant. For the excited state the argument is the same. However, now the slightly smaller momenta are closer to resonance. This combination of momentum transfer and evolution in the external potential leads to the deformation of the wave packet into an expanding banana shape.}
     \label{fig:quadratic}
\end{figure*}

\section{Conclusion}\label{sec:conclusion}
In Sec.~\ref{sec:decouple} we have shown, that a laser driven two-level system with center of mass motion in an external potential under the rotating-wave approximation can be cast into a decoupled form where everything is diagonalized with respect to the internal degrees of freedom. Depending on the potential at hand we then arrive at different driven oscillator equations for the elements of the time-evolution operator with respect to the internal states. In Sec.~\ref{sec:applications} we applied these generalized oscillators to the cases, where a potential is absent and a linear potential where we find agreement with literature and provide new, representation free analytic solutions for the time-evolution operator. Finally we have shown how the generalized oscillator can be used for a perturbative analysis of arbitrary potentials and showcase the efficiency of our approach for a quadratic potential.

The oscillation equation, Eq.~\eqref{eq:oscillator}, can be an excellent basis for the analysis of the interaction of matter and light including the external degrees of freedom of the atom. Especially, since in some cases we are able to omit the operator nature and only have to solve a classical ODE for every momentum state. This should in principle allow for highly parallel computations and can speed up numerical calculations drastically instead of split-stepping the full system.

In addition, including laser phase noise becomes straight forward once we derived the oscillation equations as only the derivative of the laser phase function enters the detuning term. 
Since the oscillator can be considered as classical for the cases of linear potentials and free motion,  a multitude of time dependent functions can be considered by drawing the analog to the time-dependent damped harmonic oscillator which is a well studied system. For many cases analytic solutions exist~\cite{Polyanin2002Oct}. One possibility is modeling laser phase noise or different pulse envelopes for the laser intensity analytically in this manner.
Moreover one has the vast and rich tools available from the study of the mathematics of damped oscillators~\cite{Chen1996Jul}.

The methods we have developed can also be used to describe a combination of state preparation and blow-away pulses~\cite{Salvi2023} and can serve as a means to model the resulting imperfections. Furthermore they can be employed to determine the phase-gradients resulting from residual sloshing of the wave-packet inside a trap during a Rabi pulse. Even beyond the applications to atomic diffraction sketched here our method is naturally suited to simplify the numerical treatment of resonantly driven spin systems in motion as it decouples spin dynamics and external motion.

\section*{Acknowledgments}
We are grateful to W. P. Schleich for his stimulating input
and continuing support. The authors are grateful to Ch.~Ufrecht, E.~Giese, G.~Janson, A.~Wolf, M.~Carmesin, E.~P.~Glasbrenner and R.~Lopp for many interesting and fruitful discussions.

We are grateful to the German Aerospace Center (Deutsche Raumfahrtagentur im Deutschen Zentrum f\"ur Luft- und Raumfahrt, DLR) for funding. The QUANTUS and INTENTAS projects are supported by the German Space Agency at the DLR with funds provided by the Federal Ministry for Economic Affairs and Climate Action (Bundesministerium f\"ur Wirtschaft und Klimaschutz, BMWK) due to an enactment of the German Bundestag under Grant Nos. 50WM2250D (QUANTUS+), 50WM2450D (QUANTUS-VI), as well as 50WM2178 (INTENTAS).

We are grateful to the Carl Zeiss Foundation (Carl-Zeiss-Stiftung) and IQST for funding in terms of the project MuMo-RmQM.

\newpage

\appendix
\section{Derivative of the Unitary $D$}\label{appendix:section:derivative}
The derivative of the exponential map is given by~\cite{Schur1891Jun,Rossmann2006Jun}
\begin{align}
    \frac{\dd}{\dd \tau} \exp{\hat{O}} &= \exp{\hat{O}}\sum\limits_{k=0}^{\infty} \frac{\qty(-1)^k}{\qty(k+1)!} \qty(\text{ad}_{\hat{O}})^k\frac{\dd}{\dd \tau} \hat{O}\\ &= \qty(\sum\limits_{k=0}^{\infty} \frac{1}{\qty(k+1)!} \qty(\text{ad}_{\hat{O}})^k\frac{\dd}{\dd \tau} \hat{O}) \exp{\hat{O}}
\end{align}
with the adjoint $\text{ad}_{\hat{O}} \hat{z} = \qty[\hat{O},\hat{z}]$.
In  Eq.~\eqref{eq:derivative_1} we have
\begin{align}
    \hat{O} = \qty(\I k\hat{z}_{\text{S}} + \I\phi(\tau))
\end{align}
with
\begin{align}
    \frac{\dd}{\dd \tau} \hat{O} &= \I\frac{\dd\phi(\tau)}{\dd\tau} + \I k \frac{\dd}{\dd \tau}\hat{z}_{\text{S}}\\
    &= \I \frac{\dd\phi(\tau)}{\dd\tau} - k \qty[\hat{T}_{\text{S}},\hat{z}_{\text{S}}].
\end{align}
Note, that the Heisenberg equation of motion enters here. The appearing commutator can be reduced to
\begin{align}
    \qty[\hat{T}_{\text{S}},\hat{z}_{\text{S}}]  &= \frac{1}{m \Omega \hbar}\hat{p}_{\text{S}} \qty[\hat{p}_{\text{S}},\hat{z}_{\text{S}}]= - \frac{\I}{m \Omega} \hat{p}_{\text{S}}.
\end{align}
Finally, we arrive at
\begin{align}
 \frac{\dd}{\dd \tau} \hat{O} =    \I \frac{\dd\phi(\tau)}{\dd\tau} + \I \frac{k}{m\Omega} \hat{p}_{\text{S}}.
\end{align}
Consequently the action of the adjoint yields
\begin{align}
\qty(\text{ad}_{\hat{O}})^0\frac{\dd}{\dd \tau} \hat{O} = \I \frac{\dd\phi(\tau)}{\dd\tau} + \I \frac{k}{m\Omega} \hat{p}_{\text{S}}\\
\end{align}
and
\begin{align}
\qty(\text{ad}_{\hat{O}})^1\frac{\dd}{\dd \tau} \hat{O} =  \qty[\I k\hat{z}_{\text{S}},\I \frac{k}{m\Omega} \hat{p}_{\text{S}}] = -\I\frac{\hbar k^2}{m\Omega}.
\end{align}
Obviously all higher orders of sequentially applying the adjoint operation yields zero, as $\qty(\text{ad}_{\hat{O}})^1\frac{\dd}{\dd \tau} \hat{O}$ is a c-number commuting with all operators. 
Therefore, we only have contributions for $k = 0$ and $k = 1$ thus the sum reduces to
\begin{align}
    \sum\limits_{k=0}^{\infty} \frac{\qty(-1)^k}{\qty(k+1)!} \qty(\text{ad}_{\hat{O}})^k\frac{\dd}{\dd \tau} \hat{O} = \I \frac{\dd\phi(\tau)}{\dd\tau} + \I \frac{k}{m\Omega} \hat{p}_{\text{S}} -\I\frac{\hbar k^2}{2m\Omega}.
\end{align} 
\section{Action of the Displacement on Operators}
For the displaced Doppler operator we find
\begin{align}   \hat{D}^{\dagger}\hat{\nu}\hat{D} = \hat{\nu} + 2 \omega_{\text{r}}\\   \hat{D}\hat{\nu}\hat{D}^{\dagger} = \hat{\nu} - 2 \omega_{\text{r}}.
\end{align}
The detuning operators fulfill the following algebraic properties under the displacement:
\begin{align}
    \hat{D}^{\dagger}\hat{\delta}_{+}\hat{D} = -\hat{\delta}_{-}\\    \hat{D}\hat{\delta}_{-}\hat{D}^{\dagger} = -\hat{\delta}_{+}.
\end{align}
\begin{widetext}
\section{Solutions for Linear Potential}\label{appendix:subsec:linear_potential}
For the linear potential, the solution of Eq.~\eqref{eq:oscillator_gravity} contains the functions~\cite{Polyanin2002Oct}
\begin{align}\label{eq:appendix_fun_A}
\begin{split}
    A\qty(\kappa_{\pm},\hat{\delta}_{\pm},\tau) &= \text{H}\qty(\frac{1}{2\I\kappa_{\pm}}-1 , \I\frac{\hat{\delta}_{\pm}+\kappa_{\pm}\tau}{\sqrt{\I\kappa_{\pm}}})
    \prescript{}{1}{\text{F}_1}\qty(\frac{2\I\kappa_{\pm}-1}{4\I\kappa_{\pm}} , \frac{1}{2} , \frac{-\hat{\delta}_{\pm}^2}{\I\kappa_{\pm}})
    \\
    &-\text{H}\qty(-\frac{\I}{2\kappa_{\pm}}-1 , \I\frac{\hat{\delta}_{\pm}}{\sqrt{\I\kappa_{\pm}}})
    \prescript{}{1}{\text{F}_1}\qty(\frac{2\I\kappa_{\pm}-1}{4\I\kappa_{\pm}} , \frac{1}{2} , \frac{-\qty(\hat{\delta}_{\pm}+\kappa_{\pm}\tau)^2}{\I\kappa_{\pm}})
\end{split}
\end{align}
\begin{align}\label{eq:appendix_fun_B}
\begin{split}
    B\qty(\kappa_{\pm},\hat{\delta}_{\pm},\tau) =& 2\sqrt{\I\kappa_{\pm}}
    \text{H}\qty(-\frac{\I}{2\kappa_{\pm}}-2 , \I\frac{\hat{\delta}_{\pm}}{\sqrt{\I\kappa_{\pm}}})
    \prescript{}{1}{\text{F}_1}\qty(\frac{2\I\kappa_{\pm}-1}{4\I\kappa_{\pm}} , \frac{1}{2} , \frac{-\hat{\delta}_{\pm}^2}{\I\kappa_{\pm}})
    \\
    &+2\I\hat{\delta}_{\pm}\text{H}\qty(-\frac{\I}{2\kappa_{\pm}}-1 , \I\frac{\hat{\delta}_{\pm}}{\sqrt{\I\kappa_{\pm}}})
    \prescript{}{1}{\text{F}_1}\qty(\frac{3}{2} - \frac{1}{4\I\kappa_{\pm}},\frac{3}{2},\frac{-\hat{\delta}_{\pm}^2}{\I\kappa_{\pm}})           
\end{split}
\end{align}
\begin{align}\label{eq:appendix_fun_C}
\begin{split}
    C\qty(\kappa_{\pm},\hat{\delta}_{\pm},\tau) &= 
    2\sqrt{\I\kappa_{\pm}}\qty(8\I\kappa_{\pm}-1)
    \text{H}\qty(-\frac{\I}{2\kappa_{\pm}}-2 , \I\frac{\hat{\delta}_{\pm}}{\sqrt{\I\kappa_{\pm}}}) \\
    &-4\hat{\delta}_{\pm}\kappa_{\pm} 
    \text{H}\qty(-\frac{\I}{2\kappa_{\pm}}-1 , \I\frac{\hat{\delta}_{\pm}}{\sqrt{\I\kappa_{\pm}}})
    \prescript{}{1}{\text{F}_1}\qty(\frac{1}{2}-\frac{1}{4\I\kappa_{\pm}} , \frac{1}{2} , \frac{-\qty(\hat{\delta}_{\pm}+\kappa_{\pm}\tau)^2}{\I\kappa_{\pm}}) \\
    &+4\kappa_{\pm}\hat{\delta}_{\pm} 
    \text{H}\qty(\frac{1}{2\I\kappa_{\pm}}-1 , \I\frac{\hat{\delta}_{\pm}+\kappa_{\pm}\tau}{\sqrt{\I\kappa_{\pm}}})
\prescript{}{1}{\text{F}_1}\qty(\frac{1}{2}-\frac{1}{4\I\kappa_{\pm}} , \frac{1}{2},\frac{-\hat{\delta}_{\pm}^2}{\I\kappa_{\pm}})\\
&- 2\I\hat{\delta}_{\pm}\qty(1-8\I\kappa_{\pm}) 
    \text{H}\qty(\frac{1}{2\I\kappa_{\pm}}-1 , \I\frac{\hat{\delta}_{\pm}+\kappa_{\pm}\tau}{\sqrt{\I\kappa_{\pm}}})
\prescript{}{1}{\text{F}_1}\qty(\frac{3}{2}-\frac{1}{4\I\kappa_{\pm}} , \frac{3}{2},\frac{-\hat{\delta}_{\pm}^2}{\I\kappa_{\pm}})
\end{split}
\end{align}
\begin{align}\label{eq:appendix_fun_G}
\begin{split}
    G\qty(\kappa_{\pm},\hat{\delta}_{\pm},\tau) &= 2
    \sqrt{\I\kappa_{\pm}} 
    \text{H}\qty(\frac{1}{2\I\kappa_{\pm}}-2 , \I\frac{\hat{\delta}_{\pm}}{\sqrt{\I\kappa_{\pm}}})
    \prescript{}{1}{\text{F}_1}\qty(\frac{1}{2}-\frac{1}{4\I\kappa_{\pm}} , \frac{1}{2} , \frac{-\hat{\delta}_{\pm}^2}{\I\kappa_{\pm}})
    \\
    &+2\I\hat{\delta}_{\pm}
    \text{H}\qty(\frac{1}{2\I\kappa_{\pm}}-1 , \I\frac{\hat{\delta}_{\pm}}{\sqrt{\I\kappa_{\pm}}})
    \prescript{}{1}{\text{F}_1}\qty(\frac{3}{2}-\frac{1}{4\I\kappa_{\pm}} , \frac{3}{2} , \frac{-\hat{\delta}_{\pm}^2}{\I\kappa_{\pm}}) 
\end{split}
\end{align}

We use the following definitions: $\text{H}\qty(a,x)$ is the Hermite Polynomial of degree $\alpha$ depending on the variable $x$ and $\prescript{}{1}{\text{F}_1}\qty(a,b, x)$ is the confluent hypergeometric function. 
\end{widetext}
\nocite{*}
 \bibliography{00_main_bib.bib}
\end{document}